# Quantum dot single-photon emission coupled into single-mode fibers with 3D printed micro-objectives

Lucas Bremer[1], Ksenia Weber[2], Sarah Fischbach[1], Simon Thiele[3], Marco Schmidt[1], Arsenty Kaganskiy[1], Sven Rodt[1], Alois Herkommer[3], Marc Sartison[4], Simone Luca Portalupi[4], Peter Michler[4], Harald Giessen[2], and Stephan Reitzenstein[1]

[1]*Institute of Solid State Physics, Technische Universität Berlin, Berlin, Germany*
[2]*4th Physics Institute and Research Center SCoPE and Integrated Quantum Science and Technology Center IQ$^{ST}$, University of Stuttgart, Stuttgart, Germany*
[3]*Institute for Applied Optics (ITO) and Research Center SCoPE, University of Stuttgart, Stuttgart, Germany*
[4]*Institut für Halbleiteroptik und Funktionelle Grenzflächen, Center for Integrated Quantum Science and Technology (IQ$^{ST}$) and Research Center SCoPE, University of Stuttgart, Stuttgart, Germany*
\* *stephan.reitzenstein@physik.tu-berlin.de*

**Abstract**

User-friendly single-photon sources with high photon-extraction efficiency are crucial building blocks for photonic quantum applications. For many of these applications, such as long-distance quantum key distribution, the use of single-mode optical fibers is mandatory, which leads to stringent requirements regarding the device design and fabrication. We report on the on-chip integration of a quantum dot microlens with a 3D-printed micro-objective in combination with a single-mode on-chip fiber coupler. The practical quantum device is realized by deterministic fabrication of the QD-microlens via in-situ electron-beam lithography and 3D two-photon laser writing of the on-chip micro-objective and fiber-holder. The QD with microlens is an efficient single-photon source, whose emission is collimated by the on-chip micro-objective. A second polymer microlens is located at the end facet of the single-mode fiber and ensures that the collimated light is efficiently coupled into the fiber core. For this purpose, the fiber is placed in the on-chip fiber chuck, which is precisely aligned to the QD-microlens thanks to the sub-µm processing accuracy of high-resolution two-photon direct laser writing. This way, we obtain a fully integrated high-quality quantum device with broadband photon extraction efficiency, a single-mode fiber-coupling efficiency of 26 %, a single-photon flux of 1.5 MHz at single-mode fibre output and a multi-photon probability of 13 % under pulsed optical excitation. In addition, the stable design of the developed fiber-coupled quantum device makes it highly attractive for integration into user-friendly plug-and-play quantum applications.

## Introduction

The development of real-world quantum communication networks [1, 2], which offer a previously unattained level of data transfer security [3, 4], has become very dynamical in recent years. In addition to the first quantum networks based on optical fibers [5, 6] and free-space channels [7, 8], also expanded solutions in the form of satellite-based quantum communication networks [9, 10] have been developed. In this context, strongly attenuated pulsed lasers are still frequently used as photon sources for decoy-state-based protocols [11, 12], offering the highest transfer rates so far. However, ultimate performance can only be obtained by using true single-photon sources, based for instance on semiconductor quantum dots (QDs), which promise not only on-demand operation but also feature nowadays the purest single-photon emission of all known quantum light sources [13, 14]. Noteworthy, only sources with non-classical single-photon statistics can exploit the full potential of quantum communication [15, 16]. In addition, the availability of true single-photon sources is crucial for the implementation of the quantum repeater protocol [17, 1]. In fact, on-demand sources of indistinguishable and entangled photons [18, 19, 20, 21] are key to make the dream of long-distance multi-partite networks become a reality. Here it is important to ensure that the photons always propagate in the same defined spatio-temporal mode [22], making the use of single-mode fibers (SMFs) as quantum channel indispensable.

On-chip fiber-coupled sources have the decisive advantage of enabling ultra-stable operation without additional free-beam optics. This feature reduces the complexity of the source enormously and, as a result boosts its flexibility and application relevance. In the past, multimode fibers [23] or SMF bundles [24] were used to realize fiber-coupled single-photon sources. Only recent progress in nanofabrication techniques has paved the way for single-mode-coupling of semiconductor QDs integrated into Fabry-Perot optical microcavities [25], cylindrical mesas [26], photonic crystal nanobeams [27] or waveguide-based devices [28, 29, 30]. In the present work we take a different approach and want to combine the strengths of QD-microlenses in terms of extraction efficiency and low multi-photon probability [31] with the advantages of precise and flexible 3D layer by layer two-photon direct laser (TPL) patterning of photonic microstructures [32, 33, 34, 35, 36].

A significant challenge to be tackled for the realization of efficient coupling of a QD-microlens to a SMF with a core diameter of only 4.4 µm and a numerical aperture (NA) as low as 0.13 is that, although QD-microlenses are mechanically very robust and provide broadband enhancement of photon extraction, their emission is only moderately directed. As a result, a micro-objective design was developed that collects and collimates the outcoupled radiation of the QD-microlens. For this purpose, the approach described in Ref. [37] was revised and extended. In the present work we developed a total internal reflection (TIR) microlens, serving as the light-collection micro-objective, which is printed with sub-micrometer accuracy onto a QD-microlens using 3D femtosecond direct laser writing. Additionally, similar to Ref. [38], we designed and realized an in-situ TPL printed on-chip fiber holder aligning the QD-microlens-system to a coupling lens on the face facet of a SMF. The synergetic combination of all these components results in a precise, fully integrated and ultra-stable micro-optical on-chip photonic system for applications in fiber-based quantum communication.

## Device Technology and Fabrication

The sample is based on a wafer heterostructure consisting of InAs QDs grown on (100) GaAs substrate by metal-organic chemical vapor deposition. A back-side distributed Bragg reflector (DBR) consisting of 23 GaAs (67 nm)/Al$_{0.9}$Ga$_{0.1}$As (78 nm) $\lambda$/4-thick layer pairs is located underneath the QD layer at a distance of 67 nm to reflect the light emitted into the lower hemisphere and, thus, increase the photon-extraction efficiency normal to the sample surface. The self-assembled QDs are randomly distributed in position and wavelength with a center wavelength of around 920 nm. Above the QD layer, a 420 nm thick capping layer is grown, which is required for deterministic structuring of QD-microlenses by 3D in-situ electron-beam lithography (EBL). With the help of numerical optimization of the microlens design it is possible to achieve outcoupling efficiencies of almost 30 % for an NA of 0.4 [31, 39].

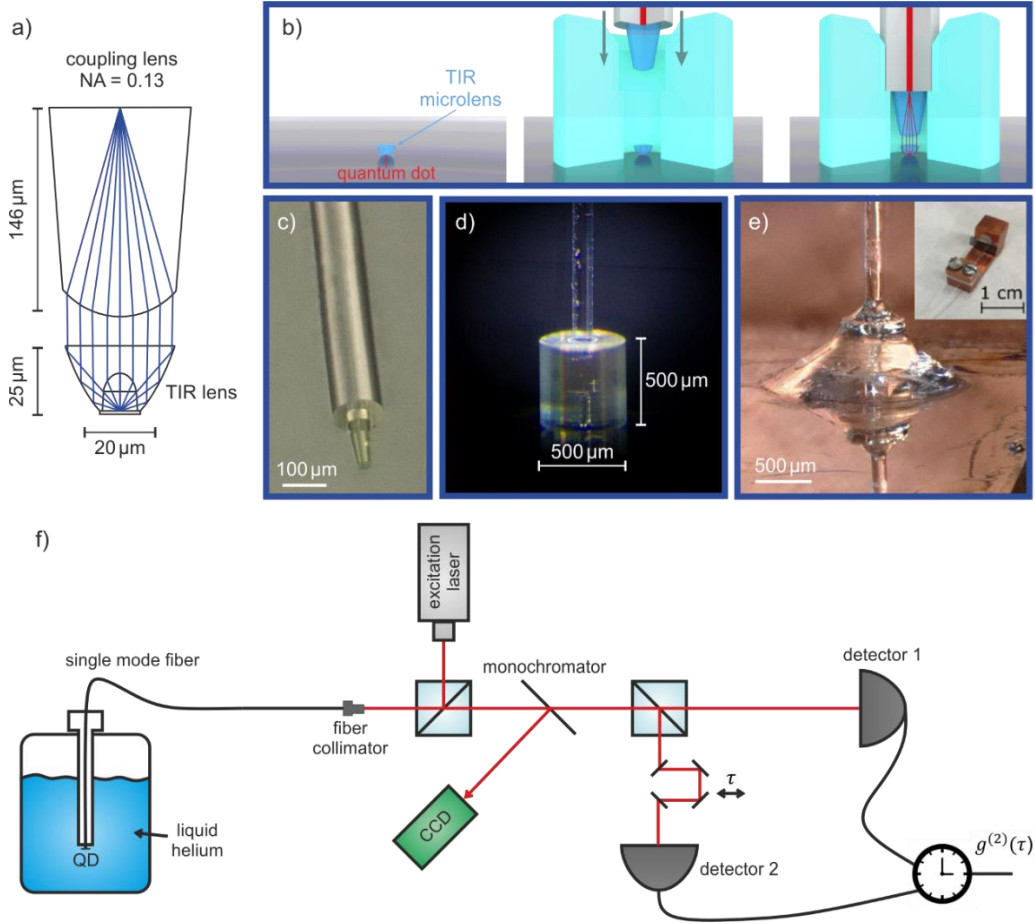

Fig. 1. (a) Optical design for coupling light from a QD-microlens into a single-mode fiber obtained from sequential ray tracing. Light that is emitted from the QD-microlens is collimated by TIR lens (NA=0.001, see supplement) and focused onto the fiber core by an NA matched (NA = 0.13) spherical focus lens. (b) Scheme showing the fabrication of an integrated fiber coupled single-photon source. A TIR lens is printed onto a QD-microlens (left), followed by 3D printing the on-chip fiber chuck and inserting the SMF with a focus lens printed to its end facet (middle). After inserting the fiber, light from the QD-microlens is efficiently coupled into the fiber core via the micro-optical system (right). (c) Microscope image of a SMF (THORLABS 780-HP) with an NA matched focus lens printed onto it. (d) Microscope image of an optical SMF inserted into a fiber holder via a manual XYZ-flexure stage. (e) Microscope image of a SMF inserted into a fiber holder and glued to the substrate with UV-cured glue. Inset: fiber-coupled QD-sample mounted onto a strain-relief copper holder for mounting in a cryostat. (f) Scheme for the second-order autocorrelation measurement. The fiber-coupled sample is inserted into a liquid helium can and excited via an off-resonant excitation laser. The fiber-coupled QD signal is then spectrally analyzed and send to a Hanbury Brown and Twiss setup with a time-delay stage in one arm to obtain the time-correlated signal.

For the fabrication of QD-microlenses, a 80 nm thick electron-sensitive CSAR62 resist film [40] is first spin-coated onto the sample and then a specially modified electron scanning microscope is used to record cathodoluminescence maps at T = 10 K. Suitable QDs are selected based on the emission wavelength and the emission intensity of the excitonic lines, and a lenticular dose profile is then introduced into the resist at the positions of selected QDs. In the later anisotropic plasma enhanced reactive ion etching step, the imposed dose profile acts as etch mask and is transferred into the semiconductor material, resulting in monolithic QD-microlenses [31].

To realize a high photon coupling efficiency into a SMF, the radiation pattern of the QD-microlens must be taken into account, i.e. to maximize the usable photon flux the emission should be collected from the largest possible solid angle and coupled into the fiber. We solve this issue by a configuration consisting of two TPL written polymer micro-optical elements as illustrated in Fig. 1(a). A total-internal reflection micro-objective on the QD-microlens is used to collimate its divergent emission, while an NA matched (NA = 0.13) coupling microlens on the single mode fiber is used to focus the beam down onto the fiber core. One advantage of this optical system is that it is rather insensitive to the distance between TIR and coupling lens, as long as the beam can still be regarded as collimated. Details of the microlens design, its optimization, and its relationship to the achievable photoluminescence enhancement are given in the supplement. For these optimization measurements, high-precision low temperature deterministic lithography was used [41] marking the emitter for further room temperature fabrication [35, 42]. Furthermore, a first characterization of the fiber in-coupling performances was carried out in a high stability cryostat where the position of the fiber with respect to the sample can be precisely controlled (see supplement). A maximum coupling efficiency of (26 +/- 2) % was observed. This motivated the fabrication of the fiber chuck to replicate this performance on a mechanically stable manner. Controlling the reciprocal position further showed that in transverse direction, precise alignment with sub-µm accuracy of all components is required [34, 33]. The micro-optics are fabricated from IP-Dip photoresist [43] with a commercial femtosecond two-photon 3D laser printer (Photonics Professional GT, Nanoscribe GmbH). The laser beam is scanned in lateral direction by two galvanometric mirrors through a 63x (NA = 1.4) microscope objective in dip-in configuration [33].

In the case of the TIR lens, this means that the photoresist is applied onto the semiconductor sample and the objective is immersed into it, while for the fiber lens, the photoresist is applied directly onto the objective and the fiber is inserted into it subsequently. In axial direction, the semiconductor sample is moved by a piezoelectric crystal, while the fiber is left stationary and the objective is moved instead.

The TIR lens is aligned to the QD using prefabricated alignment markers. As the TIR lens will cover the etched markers after its fabrication, new 3D printed markers of the same shape are printed along with the TIR lens. The SMF core is located by shining light from a red LED onto the other end facet of the fiber, which leads to the fiber core lighting up in the microscope image (more details can be found in Ref. [44]). The total printing time for the TIR lens was about 5 min. Due to its substantially larger volume, the coupling lens took about 1 h to produce. After the printing, the structures were developed in mr-Dev 600 (micro resist technology GmbH) for several minutes. During this process, unexposed photoresist is removed. After pre-characterizing QD-microlenses with TIR lenses printed onto them by µPL spectroscopy (see Fig. 2(a)), fiber holders are fabricated around the lenses using the 3D printed markers for alignment. Due to the large size of the fiber holder, the polymer IP-S which is a lower resolution photoresist and an objective with a larger field-of-view (25x, NA = 0.8) is used in this processing step. The printing time for the fiber holder is roughly 45 min. After development, the fiber integration is performed. For this, the SMF with the coupling lens printed on top (shown in Fig. 1(c)) is mounted onto a manual XYZ-flexure stage and inserted into the fiber holder under a microscope. A mirror angled at 45° is used to monitor the position of the fiber relative to the fiber holder opening in all directions.

A sketch of the fiber integration is shown in Fig. 1(b) and a microscope image of the inserted fiber in Fig. 1(d). The fiber is then fixed to the substrate by depositing a small droplet of UV-cured glue with another piece of optical fiber mounted onto a second XYZ-flexure stage. To avoid the fiber holder to detach from the substrate when cooled down to cryogenic temperatures, the UV glue is made to cover the entire fiber holder (see Fig. 1(e)). The sample is then mounted to a copper holder with a built-in strain relief as shown in the inset of Fig. 1(e). The substrate is glued to the bottom of the holder with conductive silver and the SMF is fixed about 1 cm away from it between two pieces of polytetrafluoroethylene (Teflon) with two small screws. This makes the device highly robust and allows it to be mounted in a specially designed sample holder located at the lower end of a dip stick. The dip stick is inserted into a user-friendly standard helium (He) transport vessel so that the fiber-coupled device can be studied with the setup shown in Fig. 1(f).

**Results and Discussion**

In the following we discuss the optical properties of the QD-microlens and of the complete fiber-coupled single-photon source. To quantify the fraction of QD photons coupled into the SMF, µPL spectra were recorded before and after processing. Fig. 2(a) shows a µ-PL spectrum of a QD-microlens before processing (black trace). The setup shown in Fig. 1(f) was used, except that the sample was placed in a He-flow cryostat instead of operating it directly in liquid helium. Furthermore, a microscope objective with an NA of 0.65 was used for the measurement. In the µPL spectrum the bright trion line at 916 nm is particularly noticeable besides the excitonic complex around 918 nm. The excitonic states were identified by polarization- and excitation power dependent µ-PL measurements. In addition, time-resolved measurements were used to confirm the assignment based on the decay constants of the excitonic states. Due to the highest µPL intensity the trion line was chosen for more detailed investigations in the following, including photon-autocorrelation measurements to confirm the quantum nature of emission.

Fig. 2(a) shows in red the spectrum of the same QD-microlens under comparable excitation conditions, i.e. close to saturation, after SMF coupling. Each µPL spectrum is normalized by the setup efficiency (He-flow cryostat: 5.1 %, fiber-coupled: 2.8 %) to ensure comparability of the intensities. In addition, the intensity of the fiber-coupled spectrum has been doubled to increase contrast. It is apparent that the fiber-coupling of the QD-microlens shifts the QD spectrum to lower wavelengths by a blue-shift of about 4 nm, comparable to previously observed values for 3D printed hemispheric lenses [35]. A much smaller shift of 0.2 nm due to the influence of a TPL patterned on-chip micro-objective was already discussed in Ref. [37]. In the present work, however, we use a different design for the micro-objectives which, in addition to the on-chip fiber coupling, may explain the larger blue-shift. In general, we have found that different geometries of micro-objectives consistently lead to blue shifts of 0.2 to 4.0 nm, with a tenuous correlation between the lens diameter and the resulting wavelength shift - the larger the diameter, the smaller the blue shift. The blue-shift is attributed to the compressive stress of the semiconductor material including the QD caused by the printed micro-objective. Interestingly, the wavelength shift is reversible, i.e., the QD spectrum red-shifts to the original position after removal of the micro-objective and the fiber holder (grey trace in Fig. 2(a)). This was observed when optimizing the process flow iteratively, whereby different objective geometries were printed on a variety of QD-microlenses whose emission properties were monitored after each processing step. However, the basic optical properties of the QD do not seem to be influenced by this strain effect. The linewidth of < 25 µeV is still limited by the resolution of the monochromator (25 µeV) and the lifetime (Fig. 2(c)), and the multi-photon probability (Fig. (3)) are not affected.

Besides the blue-shift, it is also noticeable that the continuous background of the spectrum is significantly reduced by more than 2/3 due to the fiber-coupling and thus, a clear and background-free spectrum can be observed from the fiber-coupled QD-microlens. This is very advantageous and is probably because the single-mode fiber facet acts effectively as a local pinhole thereby selecting emission of the target QD and suppressing contributions from the wetting layer and possibly other, non-intentionally integrated QDs centered at other positions and wavelengths. Very clean QD emission spectra are of particular interest for the use of the fiber-coupled device in stand-alone applications where for practical reasons narrow fiber-coupled optical filters are used instead of bulky and expensive monochromators [26]. These filters have usually a higher bandwidth than state of the art monochromators, which could increase multi-photon events.

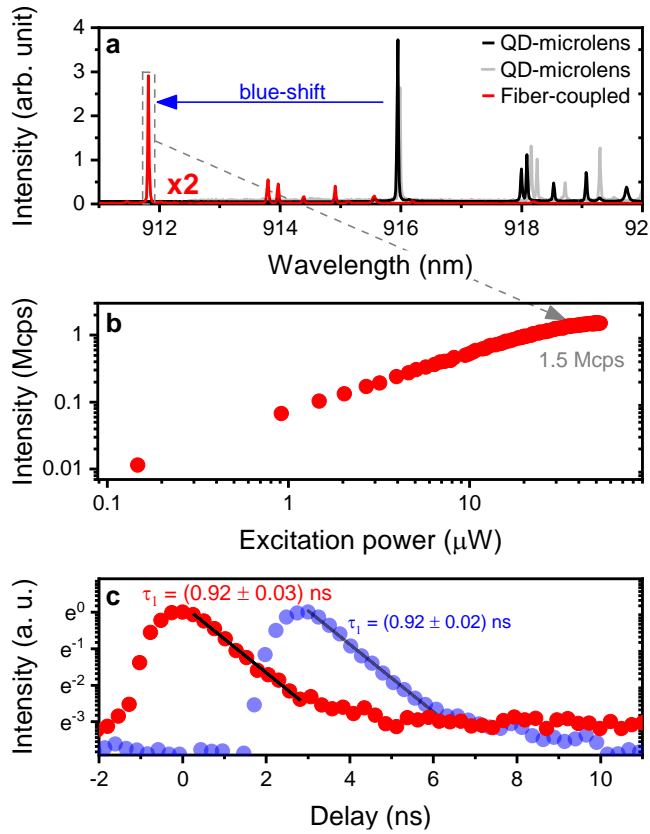

Fig. 2. (a) µPL spectrum of the QD-microlens before (black) and after (red) SMF-coupling. The induced compressive strain of the micro-objective causes a blue-shift of the QD emission lines of about 4 nm. The spectrum of the fiber-coupled sample is shown with doubled intensity for better comparability. In grey, an additional spectrum of the QD-microlens is shown, which was generated during the optimization of the micro-objective geometry. It reveals that the blue-shift is reversible by removing the micro-objective. (b) Single-photon rate of the QD emission line marked in (a) corrected by the setup efficiency in double logarithmic representation. (c) Measured lifetime for the QD line from (a) (red). The lifetime $\tau_1 = (0.92 \pm 0.03)$ ns corresponds to the inverse of the slope of a linear fit to the falling edge of the pulse. A lifetime measurement taken during the pre-characterization of the QD microlens yields $\tau_1 = (0.92 \pm 0.02)$ ns and is shown in blue (horizontally shifted for clarity). The lifetime is not influenced by the fiber-coupling.

To determine the coupling efficiency into the single-mode fiber, which is coupled and glued to the chip, we compare the integrated spectral area of the trion line before and after the processing. This comparison yields an in-coupling efficiency into the fiber of $(26 \pm 2)$ %, which reflects the excellent performance of the matched lens geometries despite the non-ideal far-field pattern of the QD-microlens. If the trion line of the spectrum shown in Fig. 2(a) is spectrally selected with a monochromator and detected with an avalanche-photodiode-based single-photon counting module (SPCM), we achieve a count rate of 42 kHz at saturation. Considering the overall transmission of our setup of 2.8 %, this corresponds to a single-photon photon flux of 1.5 MHz at the end of the single-mode fiber. Fig. 2(b) indicates the count rate also for lower excitation powers. Fig. 2(c) depicts in red a time-resolved measurement of the lifetime of the trion-transition of the fiber-coupled sample, which can be fitted using an exponential decay function to quantify the decay time $\tau_1 = (0.92 \pm 0.03)$ ns. The result is in quantitative agreement with the value ($\tau_1 = (0.92 \pm 0.02)$ ns) recorded on the free-space configuration before fiber-coupling and shows that fiber-coupling does not influence the lifetime of the QD-microlens.

The emission of the trion line is spectrally filtered by a monochromator and coupled at the monochromator exit slit into a fiber-based HBT setup for coincidence measurements (see Fig. 1(f)). The off-resonant excitation power under cw excitation was chosen to saturate the trion line (see Fig. 2(b)). In Fig. 3(a) the measured second-order photon autocorrelation function is plotted showing clear antibunching at $\tau = 0$ ns. A physical description of the measurement is possible by a two-sided exponential function

$$g^{(2)}(\tau) = 1 - \left( \left(1 - g^{(2)}(0)\right) e^{-\frac{|\tau|}{\tau_{sp}}} \right). \qquad (1)$$

The fitting of the equation to the measurement data yields $g^{(2)}(0) = 0.00^{+0.04}_{-0.00}$ which verifies a very low probability of multi-photon emission events. Noteworthy, the spontaneous decay time of $\tau_{sp} = (0.93 \pm 0.05)$ ns resulting from this fit is in very good agreement with the lifetime measurement shown in Fig. 2(c).

Furthermore, the quantum nature of emission was also characterized under the more application relevant pulsed excitation. For this purpose, the wavelength of a tunable titanium-sapphire laser was set to 865 nm, i.e. resonant with the wetting layer, and the excitation power was reduced until the detected count rate of the trion line had halved to reduce the contribution of uncorrelated background emission which increases in relative strength with the excitation power. Fig. 3(b) displays the corresponding correlation histogram.

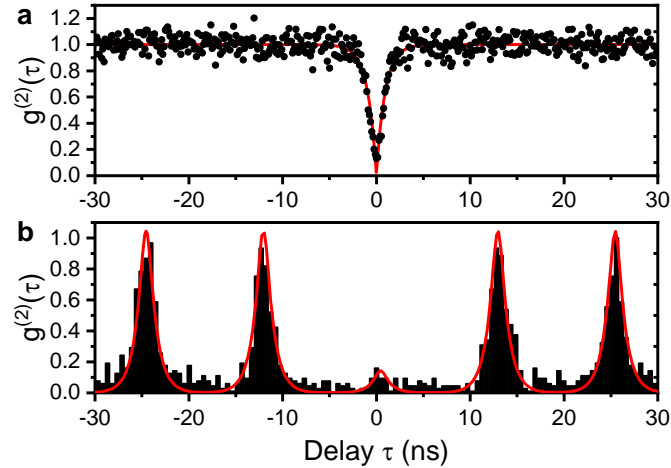

Fig. 3. Measured second-order autocorrelation function. The histograms show single-photon emission with low multi-photon probability of the fiber-coupled device. (a) Normalized histogram for non-resonant cw excitation. The red solid line shows an exponential fit of the measured data without consideration of the time resolution of the setup (without deconvolution). This results in a value of: $g^{(2)}(0) = 0.00^{+0.04}_{-0.00}$. (b) Photon autocorrelation histogram for non-resonant pulsed excitation at 80 MHz. The red line represents a fit to the data, assuming a mono-exponential radiative decay of $\tau_1 = 0.92$ ns and considering the overall time resolution of the HBT setup of 495 ps. In this approach one obtains a value of $g^{(2)}(0) = 0.13 \pm 0.05$.

To determine the $g^{(2)}(0)$-value, we fitted the experimental data with a sequence of equidistant photon pulses, each represented by the convolution of a two-sided exponential decay ($\tau_1 = 0.92$ ns) with a Gaussian function of 495 ps width (full width at half maximum), considering the time resolution of the HBT setup. Assuming a constant area A of the finite-time-delay pulses, the $g^{(2)}(0)$-value is expressed by the ratio $A_0/A$, where $A_0$ corresponds to the area of the zero-time-delay peak. This evaluation results in $g^{(2)}(0) = 0.13 \pm 0.05$.

The differences in the $g^{(2)}(0)$-values under cw and pulsed excitation can presumably be attributed to charge-carrier recapture processes in the case of pulsed excitation [45] which can no longer be neglected and, in addition, the autocorrelation of the laser using the HBT-setup indicated that laser pulses of lower intensity were also observed outside the expected time windows given by the repetition rate. Nevertheless, the device fulfils the requirements of a fiber-coupled pure single photon source on demand, whereby the slight increase of multi-photon events in the case of pulsed excitation compared to cw excitation could certainly be improved by resonant excitation schemes.

**Conclusion**

In summary, we realized a user-friendly single-mode fiber-coupled single-photon source with excellent optical and quantum optical properties. The device fabrication benefits from the synergetic combination of 3D electron-beam lithography and femtosecond 3D direct laser writing, which enables us to join the advantages of both methods and to realize a robust, single-mode fiber coupled micro-optical on-chip system based on a single pre-selected semiconductor QD. Explicitly, a QD-microlens produced using a powerful 3D in-situ technique was combined with a total-internal-reflection micro-objective using 3D direct laser writing. The high control achievable on the printed lens allows for a precise optimization of the emitter far field. In addition, a fiber holder was written with sub-µm alignment accuracy onto the deterministically fabricated QD-microlens, which allows a SMF with a 3D printed incoupling lens on its facet to be inserted and permanently attached to the QD-microlens – micro-objective assembly. A coupling efficiency of 26 % was determined for the fiber-coupled source, leading to a maximum single-photon rate of 1.5 Mcounts/second at the output of the fiber. This technology concept has high potential to pave the way for single-mode coupled stand-alone single-photon sources of high emission rate and quantum optical quality in future, as demanded with regard to the implementation of scalable quantum networks.

**Acknowledgments**

S.R. acknowledges funding from the German Federal Ministry of Education and Research (BMBF) through the project Q.Link.X, the German Research Foundation via the project Re2974/10-1, and from the projects EMPIR 14IND05 MIQC2 and SIQUST (the EMPIR initiative is co-funded by the European Union's Horizon 2020 research and innovation program and the EMPIR Participating States). H.G. and K.W. acknowledge funding by ERC (AdG ComplexPlas, PoC 3DPrintedoptics), BMBF (Q.Link.X, Printoptics, Printfunction), MWK BW (ZAQuant), BW Stiftung (Opterial) and DFG (SPP1839 and 1929). We thank the Physikalisch-Technische Bundesanstalt Berlin for technical support. S.L.P, M.S. and P.M. greatly acknowledge funding from the BMBF (German Federal Ministry of Education and Research) via the project Q.Link.X (No. 16KIS0862) and support via the project EMPIR 17FUN06 SIQUST. This project received funding from the EMPIR programme cofinanced by the Participating States and from the European Union's Horizon 2020 research and innovation program.

**Data availability**
The data that support the findings of this study are available from the corresponding author upon reasonable request.

**Supplementary Material**

**Quantum dot single-photon emission coupled into single-mode fibers with 3D printed micro-objectives**

Lucas Bremer[1], Ksenia Weber[2], Sarah Fischbach[1], Simon Thiele[3], Marco Schmidt[1], Arsenty Kakganskiy[1], Sven Rodt[1], Alois Herkommer[3], Marc Sartison[4], Simone Luca Portalupi[4], Peter Michler[4], Harald Giessen[2], and Stephan Reitzenstein[1]

[1]Institute of Solid State Physics, Technische Universität Berlin, Berlin, Germany

[2]4th Physics Institute and Research Center SCoPE and Integrated Quantum Science and Technology Center IQST, University of Stuttgart, Stuttgart, Germany

[3]Institute for Applied Optics (ITO) and Research Center SCoPE, University of Stuttgart, Stuttgart, Germany

[4]Institut für Halbleiteroptik und Funktionelle Grenzflächen, Center for Integrated Quantum Science and Technology (IQST) and Research Center SCoPE, University of Stuttgart, Stuttgart, Germany

*stephan.reitzenstein@physik.tu-berlin.de

In this supplemental section, we would like to briefly discuss the choice of solid immersion lens shape on top of the quantum dots and its implications for the collection efficiency enhancement.

3D printed nano-objects are a very interesting approach for reliable and flexible design of micro-optics with the scope of increasing the light extraction from solid-state emitters as well as for manipulating the emission far-field. Indeed, even if the extraction efficiency enhancement is limited by the achievable refractive index mismatch between the solid-state environment (here ~3.5 at ~900 nm at cryogenic temperatures) and the utilized resist (here ~1.5) high flexibility in terms of lens geometry can be achieved.

In the current study, we utilized a geometry called total internal reflection solid immersion lenses (TIR-SIL). The observed performances allow for validating the use of deterministically placed micro objects to enhance the light extraction even for small numerical apertures. The sensitivity under lateral displacement between emitter and 3D printed object is here circumvented by using deterministic lithography techniques [S1, S2], having demonstrated the possibility to interface these techniques with 3D printing [S3, S4]. For the study on TIR-SIL a relatively simple sample geometry is used, i.e. a QD layer grown on top of a single distributed Bragg reflector, with a thick 130 nm GaAs top cap. Thanks to the deterministic lithography, a precise quantification of the extraction efficiency by direct comparison of the spectra in saturation before and after the 3D printing can be achieved [S1]. In order to ensure that the optical setup (standard micro-photoluminescence in free space, with a sample mounted in a cold finger cryostat) is consistently aligned, reference QDs are deterministically preselected and marked, representing a reference signal in different measurement runs.

The first validation measurements are therefore performed by deterministically fabricating TIR-SILs on preselected QDs. In this structure, total internal reflection between lens and air is used to fold the extracted light into small apertures. In the current study, a TIR-SIL geometry has been utilized, where the folding NA is set to 0.001. A measured extraction efficiency enhancement of 7.4 +/- 0.49 was observed and the small deviation from theoretical values (11.96) can also be attributed to small deviation from ideal design.

These results motivated the use of a fiber for the light coupling. For this purpose, a collection lens was 3D printed on the fiber tip to focus the light emitted from the QD in the single mode of the fiber. Using a high stability bath cryostat, the fiber was aligned with respect to the sample, demonstrating that (26 +/- 2)% of the emitted light can be coupled in the fiber.

Having validated the use of 3D printed optics to extract more light and manipulate the beam far-field, the brightest available sample was used for the successive step. For the final structure, i.e. deterministically fabricated QD-microlens, with 3D printed TIR-SIL and fiber chuck, a different sample has been used.

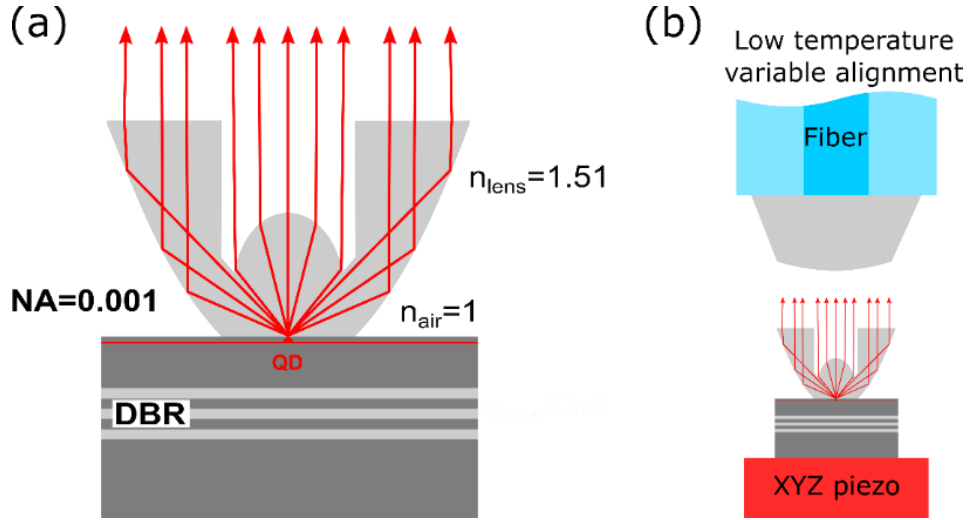

Fig. S1. (a) Sketch of the unitized TIR-SIL, with folding NA of 0.001. The bottom DBR (40 pairs) is schematically shown. (b) Validation measurement: the sample is mounted on a high precision xyz piezo stack (nanometric movement accuracy). Its position can be varied as a function of the collection fiber.

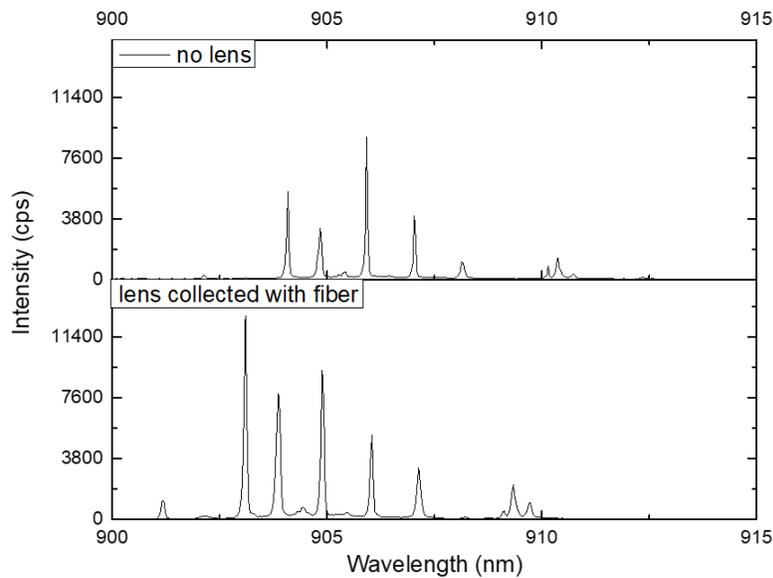

Fig. S2. (top) Spectrum of the selected QD before the lens printing. (bottom) Fiber coupled signal after the TIR-SIL printing (fig. S1 (b)). Analysis as in Ref. S2 yield a coupling efficiency of (26 +/- 2)%. Typical spectral shift of ~ 1 nm is observed. Spectra are in saturation under above band gap pumping (~ 660 nm).